\documentclass{PoS}

\usepackage{amssymb,amsmath,graphicx,epsfig,bm,color}
\usepackage[caption=false]{subfig}

\renewcommand{\d}{\mathrm{d}}
\newcommand{\be}{\begin{equation}}
\newcommand{\ee}{\end{equation}}
\newcommand{\bea}{\begin{eqnarray}}
\newcommand{\eea}{\end{eqnarray}}

\newcommand{\pup}{p^\uparrow}

\def\lsim{\mathrel{\rlap{\lower4pt\hbox{\hskip1pt$\sim$}}\raise1pt\hbox{$<$}}}
\def\gsim{\mathrel{\rlap{\lower4pt\hbox{\hskip1pt$\sim$}}\raise1pt\hbox{$>$}}}
\def\nostrocostruttino#1\over#2{\mathrel{\mathop{\kern 0pt \rlap
{\hbox{$#1$}}} \hbox{\kern-.135em $#2$}}}

\title{Process dependence of the gluon Sivers function in inclusive $\bm{pp}$ collisions: theory}

\ShortTitle{Process dependence of the gluon Sivers function in inclusive $pp$ collisions: theory}

\author{\speaker{Cristian Pisano}\\ 
        Dipartimento di Fisica, Universit\`a di Cagliari and INFN, Sezione di Cagliari \\
        Cittadella Universitaria, I-09042 Monserrato (CA), Italy\\
        E-mail: \email{cristian.pisano@ca.infn.it}}

\author{Umberto D'Alesio\\
        Dipartimento di Fisica, Universit\`a di Cagliari and INFN, Sezione di Cagliari \\
        Cittadella Universitaria, I-09042 Monserrato (CA), Italy\\
        E-mail: \email{umberto.dalesio@ca.infn.it}}

\author{Carlo Flore\\
        Dipartimento di Fisica, Universit\`a di Cagliari and INFN, Sezione di Cagliari \\
        Cittadella Universitaria, I-09042 Monserrato (CA), Italy\\
        E-mail: \email{carlo.flore@ca.infn.it}}

\author{Francesco Murgia\\
        INFN, Sezione di Cagliari, 
        Cittadella Universitaria, I-09042 Monserrato (CA), Italy\\
        E-mail: \email{francesco.murgia@ca.infn.it}}

\author{Pieter Taels\\
        INFN, Sezione di Cagliari, 
        Cittadella Universitaria, I-09042 Monserrato (CA), Italy\\
        E-mail: \email{pieter.taels@ca.infn.it}}

\abstract{Within the color gauge invariant generalized parton model (CGI-GPM), which includes initial and final state interactions in a transverse momentum dependent formalism, we present the complete results, at leading order in perturbative QCD, for transverse single-spin asymmetries in the inclusive hadroproduction of $J/\psi$ and $D$ mesons, pions and direct photons.}

\FullConference{23rd International Spin Physics Symposium - SPIN2018 -\\
		10-14 September, 2018\\
		Ferrara, Italy}

\begin{document}

\section{Introduction and formalism}
\label{theory}

Transverse single-spin asymmetries (SSAs) in high-energy lepton-proton and proton-proton collisions can provide important information on the three-dimensional structure of the proton. For this reason, they have stimulated the research programs of several experimental collaborations, like HERMES at DESY in the past and, more recently, COMPASS at CERN, CLAS(12) at Jefferson Lab and BRAHMS, PHENIX and STAR at RHIC.  

From the theoretical point of view, SSAs  for the inclusive production of hadrons and photons in $pp$ scattering, namely $p^\uparrow p \to h\, X$ and $p^\uparrow p \to \gamma\,  X$,  are defined as 
\begin{equation}
A_N \equiv \frac{\d\sigma^\uparrow-\d\sigma^\downarrow}{\d\sigma^\uparrow+\d\sigma^\downarrow} \equiv\, \frac{ \d\Delta\sigma}{ 2 \d\sigma}\,,
\end{equation}
where $\d\sigma^{\uparrow (\downarrow)}$ denotes the cross section in which one of the protons is transversely polarized with respect to the production plane along a direction $\uparrow$ ($\downarrow$).  
Within the framework of the so-called generalized parton model (GPM)~\cite{DAlesio:2007bjf}, which includes both spin and transverse momentum effects, the numerator of $A_N$ is sensitive to the quantity
\bea
\Delta \hat f_{a/\pup}\,(x_a, \bm k_{\perp a}) &\equiv&
\hat f_{a/\pup}\,(x_a, \bm k_{\perp a}) - \hat f_{a/p^\downarrow}\,
(x_a, \bm k_{\perp a})
=  -2 \, \frac{k_{\perp a}}{M_p} \, f_{1T}^{\perp a} (x_a, k_{\perp a}) \>
\cos\phi_a \, ,
\label{defsiv}
\eea
where $\hat f_{a/\pup}\,(x_a, \bm k_{\perp a})$ is the number density of partons $a$ with light-cone momentum fraction $x_a$ and transverse momentum $\bm k_{\perp a} = k_{\perp a} (\cos\phi_a, \sin\phi_a)$ inside the transversely polarized proton, which is assumed to move along the $\hat z$-axis. Furthermore, $M_p$ is the proton mass, while the Sivers function  $f_{1T}^{\perp a} (x_a, k_{\perp a})$~\cite{Sivers:1989cc} describes the azimuthal distribution of the unpolarized parton $a$ inside the transversely polarized proton and  fulfills the positivity bound
\begin{equation}
2\, \frac{k_{\perp a}}{M_p}\, \vert f_{1T}^{\perp a} (x_a, k_{\perp a})\vert \le   \hat f_{a/\pup}\,(x_a, \bm k_{\perp a}) \,+\,\hat f_{a/p^\downarrow} \, (x_a, \bm k_{\perp a})  \, \equiv 2\,  f_{a/p}\,(x_a, \bm k_{\perp a})\,,
\label{eq:posbound}
\end{equation}
with $f_{a/p}\,(x_a, \bm k_{\perp a})$ being the transverse momentum dependent distribution (TMD) of an unpolarized parton $a$ inside an unpolarized proton.  The Sivers function is sometimes also denoted as  $\Delta^N f_{a/\pup} \equiv-(2 \, {k_{\perp a}}/{M_p} )\, f_{1T}^{\perp a}$. 


In the GPM approach, the Sivers function is assumed to be universal. However, universality is expected to be broken by the presence of non-perturbative initial (ISI) and final (FSI) state interactions of the active partons with the polarized proton remnants.  In the Color Gauge Invariant formulation of the GPM, named CGI-GPM, the effects of ISI and FSI  have been initially taken into account  only in  the calculation of the {\em quark} Sivers function~\cite{Gamberg:2010tj,D'Alesio:2011mc}, by adopting a one-gluon exchange approximation.  It turns out that the quark Sivers function can still be taken to be universal, but needs to be convoluted with modified partonic cross sections, into which the process dependence is absorbed.  In terms of Mandelstam variables, these cross sections are the same as in  the twist-three collinear approach~\cite{Gamberg:2010tj}. We note that the CGI-GPM  is able to recover the expected opposite relative sign of the Sivers functions for Semi-Inclusive Deep Inelastic Scattering (SIDIS) and the Drell-Yan (DY) processes~\cite{Collins:2002kn,Brodsky:2002rv}.

When extending this model to the gluon sector~\cite{DAlesio:2017rzj,DAlesio:2018rnv}, one has to introduce two independent gluon Sivers functions, since for three colored gluons there are two different ways of forming a color singlet state, {\it i.e.}\ through an $f$-type (totally antisymmetric, even under charge conjugation) and a  $d$-type  (symmetric and odd under $C$-parity) color combination.  Hence, along the lines of Ref.~\cite{Bomhof:2006ra}, we introduce an $f$-type and a $d$-type gluon Sivers distribution. 

\begin{figure}[t]
\begin{center}
\includegraphics[trim= 60 710 60 45,clip,width=15.2cm]{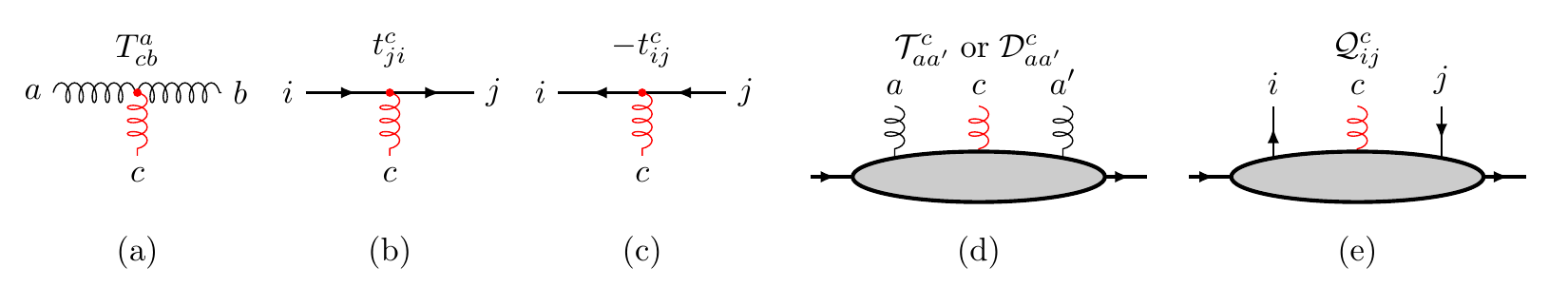}
\end{center}
\caption{CGI-GPM color rules for the following vertices involving an eikonal gluon with color index $c$: three gluons (a),  quark-gluon (b) and  antiquark-gluon (c). The two color projectors for the gluon Sivers function (d), and the one for the quark Sivers function (e), are presented as well.}
\label{fig:Feynman}
\end{figure}

\section{Single spin asymmetries in the CGI-GPM framework}

In this section, we provide the explicit expressions of the numerators of the asymmetries, valid at leading order (LO) in perturbative QCD, for the processes $p^\uparrow p \to J/\psi\, X$, $p^\uparrow p \to D\, X$,  $p^\uparrow p \to \pi\, X$ and $p^\uparrow p \to \gamma\, X$, which can be used to gather information on the so-far poorly known gluon Sivers distribution. Schematically, in the GPM,  we can write
\begin{align}
 \d\Delta\sigma \, \propto\, \sum_{a,b,c,d}\, \left  (- \frac{k_{\perp\,a}}{M_p}\right )  f_{1T}^{\perp\,a}(x_a,  k_{\perp a}) \cos\phi_a\otimes 
 f_{b/p}(x_b, \bm k_{\perp b}) \otimes H_{ab\to c d }^U \otimes D_{h/c}(z,\bm k_{\perp h})\,,
\label{eq:num-SSA}
\end{align}
where  $\otimes$ denotes a convolution in the light-cone momentum fractions and transverse momenta. Furthermore,  $z$ is the light-cone momentum fraction of the fragmenting parton $c$ carried by hadron $h$, and $\bm k_{\perp h}$ is the transverse momentum of $h$ w.r.t.\ the direction of $c$. For photon production, $D_{h/c}$ is replaced by $\delta(1-z)\delta^2(\bm k_{\perp h})$. The hard functions $H^U_{ab\to c d }$ are related to the usual unpolarized cross sections for the partonic subprocesses $ab\to cd$ and can be calculated perturbatively.

Formally, the numerator of the asymmetry in the CGI-GPM  can be obtained from Eq.~(\ref{eq:num-SSA}) by performing the following substitution
\begin{align}
f_{1T}^{\perp\,q} \otimes H_{qb\to c d}^U   \longrightarrow f_{1T}^{\perp\,q}\otimes H_{qb\to cd}^{{\rm Inc}}\,,
\end{align}
when   $a=q$, {\it i.e.}\ when parton $a$ inside the polarized proton is a quark, and similarly when $a$ is an antiquark. The modified partonic hard functions are denoted by  $H_{qb\to cd}^{{\rm Inc}}$ and can be obtained by inserting an extra eikonal gluon in the LO Feynman diagrams for the process under study and applying the rules in Fig.~\ref{fig:Feynman}. The color projector for the quark Sivers function in Fig.~\ref{fig:Feynman}~(e) is 
 \begin{equation}
 {\cal Q}^c_{ij}= {\cal N}_{\cal Q} \, t^c_{ij} \,,\qquad  \text{with}\quad {\cal N}_{\cal Q} = \frac{1}{ \text{Tr}[t^ct^c]} = \frac{2}{N_c^2-1}\,,
 \end{equation}
where $t^c_{ij}$  are the generators of the group $SU(N_c)$  in the fundamental representation and $N_c$ is the number of colors. 

When $a=g$, we have to consider the following two contributions   
\begin{align}
f_{1T}^{\perp\,g} \otimes H_{gb\to c d}^U   \longrightarrow f_{1T}^{\perp\,g\,(f)}\otimes H_{gb\to cd}^{{\rm Inc} \,(f)}\, \, + \,\, f_{1T}^{\perp\,g\,(d)} \otimes H_{gb\to cd}^{{\rm Inc}\,(d)}\,,
\end{align}
corresponding to the two different gluon Sivers functions. The $f$- and $d$-type modified hard functions can be calculated using the color projectors in Fig.~\ref{fig:Feynman}~(d), namely
\begin{equation}
{\cal T}^{c}_{a a^\prime}   =  {\cal N}_{\cal T}\, T^c_{a a^\prime}\,, \qquad \qquad {\cal D}^{c}_{a a^\prime}   =  {\cal N}_{\cal D}\, D^c_{a a^\prime}\, ,
\label{eq:cp}
\end{equation}
with  $T^c_{a a^\prime} \equiv- i f_{a a^\prime c}$ and $D^c_{a a^\prime}\equiv d_{aa^\prime c}$ being the antisymmetric and symmetric structure constants of $SU(N_c)$, respectively. The corresponding normalization factors read
\begin{equation}
 {\cal N}_{\cal T} = \frac{1}{{\rm Tr}[T^cT^c]} = \frac{1}{N_c(N_c^2-1)}\,,\qquad \qquad  {\cal N}_{\cal D} = \frac{1}{{\rm Tr}[D^cD^c]} = \frac{N_c}{(N_c^2-4)(N_c^2-1)}\,.
 \label{eq:norm-cp}
\end{equation}

In the following, we show the derivation of $A_N$ for $pp^\uparrow \to J/\psi\,X$ in the CGI-GPM approach and present  the results for the other aforementioned processes. 


\begin{figure}[t]
\begin{center}
\hspace*{0.2cm}
\subfloat[]{\hspace*{-0.3cm}\includegraphics[trim= 203 600 120 50,clip,width=5.2cm]{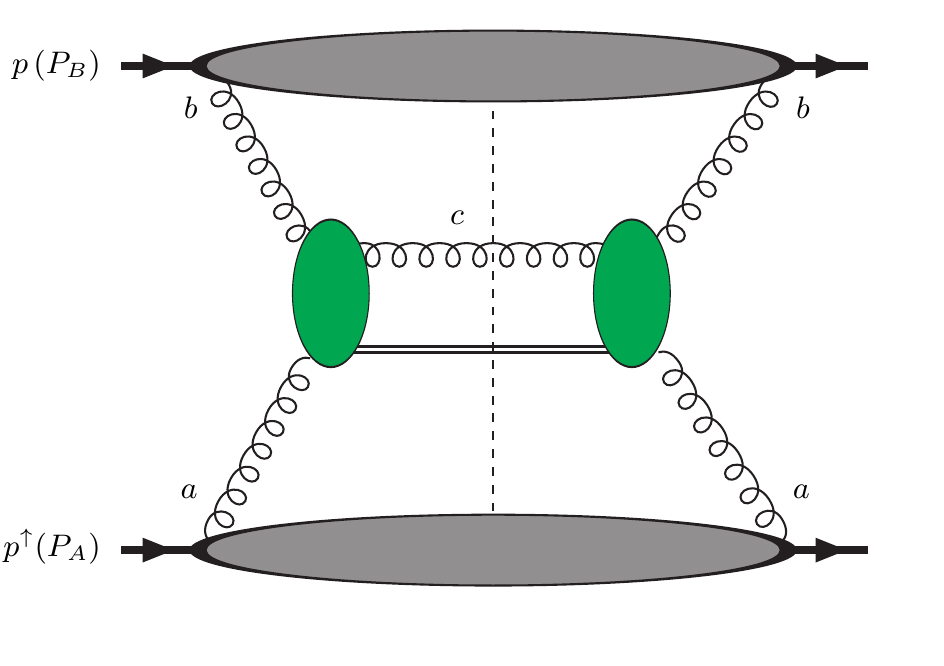}}
\hspace*{0.1cm}
\subfloat[]{\hspace*{-0.3cm}\includegraphics[trim= 203 600 120 50,clip,width=5.2cm]{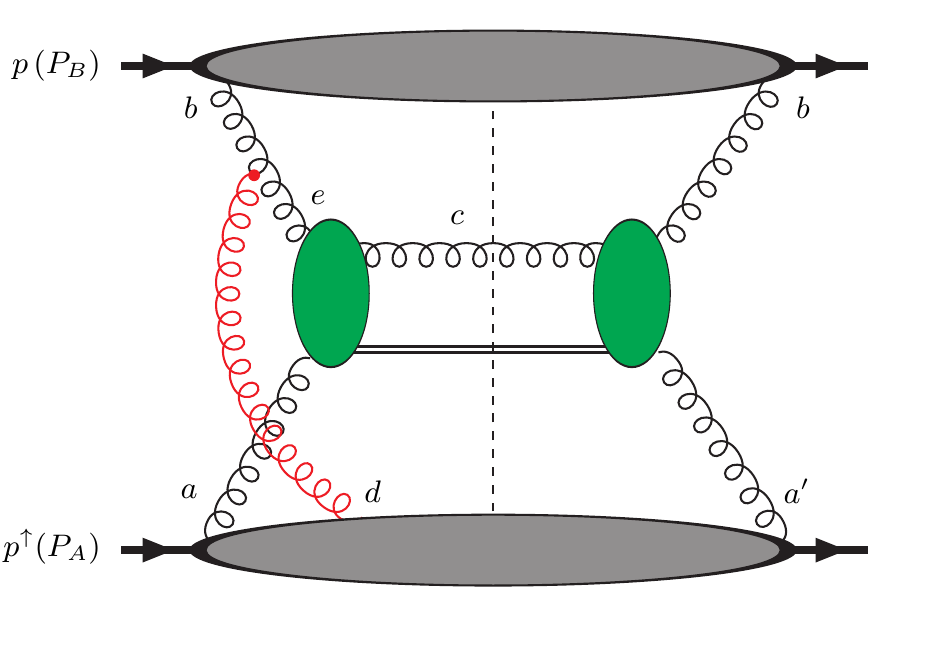}}
\hspace*{0.1cm}
\subfloat[]{\hspace*{-0.3cm}\includegraphics[trim= 203 600 120 50,clip,width=5.2cm]{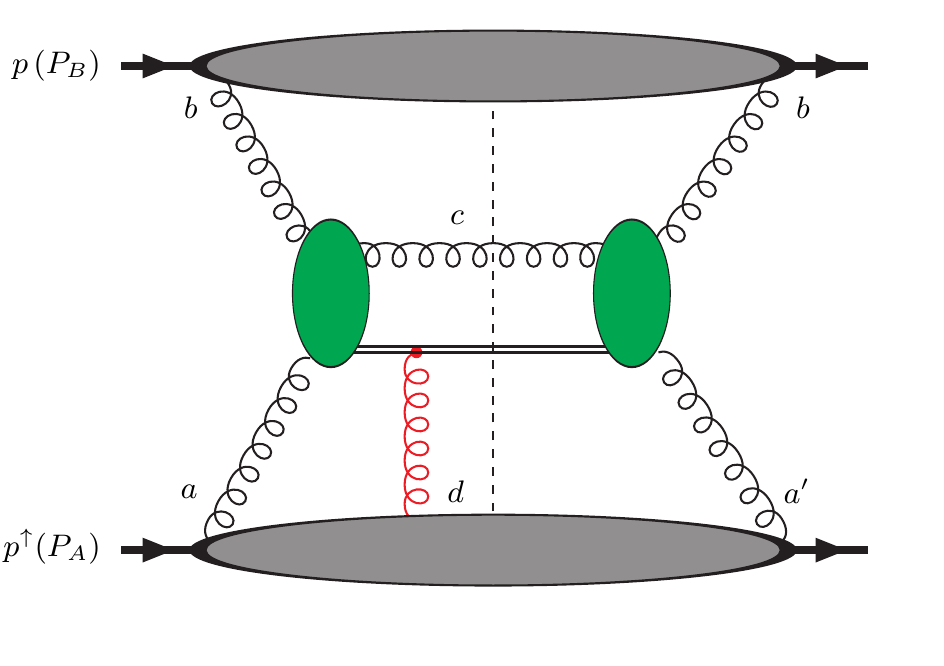}}
\end{center}
\caption{LO diagrams for the process $p^\uparrow p \to J/\psi\,X$, assuming a color-singlet production mechanism, within the GPM (a) and the CGI-GPM (b), (c). It turns out that only initial state interactions depicted in (b) contribute to the SSA.}
\label{fig:gglo-CGI}
\end{figure}

\subsection{$p^\uparrow \,p \,\to \,J/\psi\,X$}

For RHIC kinematics and small values of the transverse momentum of the $J/\psi$, the dominant production mechanism should be the Color Singlet one, according to which the charm-anticharm pair forming the bound state is produced in the hard partonic scattering directly with the same quantum numbers as the $J/\psi$. Therefore, at leading order, $gg \to J/\psi\, g$ is the main underlying partonic subprocess, with the additional real gluon emission in the final state required by the Landau-Yang theorem.  Hence, in the GPM, the numerator of the asymmetry is given by Eq.~\eqref{eq:num-SSA} with $D_{h/c} (z, \bm k_{\perp h} )\to \delta(1-z)\delta^2(\bm k_{\perp \pi})$ and
\begin{equation}
H^U_{gg\to J/ \psi g}  = \frac{5}{9}\, \vert R_0(0)\vert^2 \, M\,
\frac{\hat s^2 (\hat s-M^2)^2 + \hat t^2 (\hat t-M^2)^2 + \hat u^2 (\hat u -M^2)^2 }
{(\hat s -M^2)^2 (\hat t -M^2)^2 (\hat u -M^2)^2}\,,
\label{eq:HU}
\end{equation}
as can be obtained from the diagram in Fig.~\ref{fig:gglo-CGI}~(a). In particular, the color factor is given by
\begin{equation}
C_U= \frac{1}{(N_c^2-1)^2}\, \frac{1}{4 N_c}\, D^a_{bc}\, D^a_{cb} = \frac{1}{(N_c^2-1)^2}\, \frac{1}{4 N_c}\, \left [ \frac{(N_c^2-4)(N_c^2-1)}{N_c} \right ] = \frac{N_c^2-4}{4 N_c^2(N_c^2-1)} \,,
\label{eq:CU}
\end{equation}
{\it i.e.}\ $C_U = 5/288$ for $N_c=3$. In Eq.~\eqref{eq:HU},  $R_0(0)$ is the value of the $J/\psi$ radial wave function calculated at the origin, $M$ is the $J/\psi$ mass and $\hat s$, $\hat t$, $\hat u$ are the usual Mandelstam variables for the partonic process $gg \to J/\psi\, g$. 

In the CGI-GPM the effects of the ISI and FSI are described by the insertion of a longitudinally polarized gluon $A^+$ with momentum $k^\mu\approx k^+$ and color index $d$, as shown in Figs.~\ref{fig:gglo-CGI}~(b), (c).  It is the imaginary part of the new eikonal propagators that provides the phase needed to generate the Sivers asymmetry. As shown in Ref.~\cite{DAlesio:2017rzj},  the modified partonic hard functions turn out to be
\begin{align}
H_{gg\to J/\psi g}^{{\rm Inc} \,(f/d)} =   \frac{C_I^{(f/d)} + C^{(f/d)}_{F_c}}{C_U}\,  H_{gg\to J/\psi g}^U~.
\label{eq:mod-H}
\end{align}
The color factors $C_I^{(f/d)}$ are calculated from Fig.~\ref{fig:gglo-CGI}~(b) as follows
\begin{align}
C_{I}^{(f)} & =\frac{1}{N_{c}^{2}-1}{\cal T}_{aa^{\prime}}^{d}T_{eb}^{d} \,\frac{1}{4 {N_c}}\, D^e_{ac} D^{b}_{ca'}  = - \frac{1}{4 N_c^2(N_c^2-1)^2}\,  T^d_{a^\prime a} D^{c}_{a e} T^d_{eb}D^c_{ba^{\prime}} =-\frac{1}{2}C_{U},\label{eq:CI}
\end{align}
where we have used Eqs.~\eqref{eq:cp}, \eqref{eq:norm-cp} and the identity  ${\rm Tr}\left [ T^dD^cT^dD^c \right]  =  (N_c^2-1) (N_c^2-4)/2$.  Analogously, for the $d$-type color factor we find
\begin{align}
C_{I}^{(d)} & = \frac{1}{N_{c}^{2}-1}{\cal D}_{aa^{\prime}}^{d}T_{eb}^{d} \,\frac{1}{ 4 {N_c}}\, D^e_{ac} D^{b}_{ca'} = 0
\end{align}
because ${\rm Tr}\left [ D^dD^cT^dD^c \right]  = 0$. As already pointed out in Ref.~\cite{Yuan:2008vn}, the net contribution of the heavy charm-anticharm pair to the FSI  in Fig.~\ref{fig:gglo-CGI}~(c) is zero because the pair is produced in a color singlet state. Therefore we get
\begin{align}
C_{F_c}^{(f)} & =  C_{F_c}^{(d)} = 0\,,
\end{align}
which implies
\begin{equation}
H^{{\rm Inc} \,(f)}_{gg\to J/\psi g}  = -\frac{1}{2}\, H^{U}_{gg\to J/\psi g}\,, \qquad H^{{\rm Inc} \,(d)}_{gg\to J/\psi g}  =0.
\end{equation}
Notice that we did not consider the FSI of the unobserved particle (the gluon with color index $c$ in Fig.~\ref{fig:gglo-CGI}) because they are known to vanish after summing the different cut diagrams~\cite{Gamberg:2010tj}. 

We conclude that this process can be very useful to obtain direct information on $f_{1T}^{\perp\,g\,(f)}$ because $f_{1T}^{\perp\,g\,(d)}$ does not contribute to the asymmetry.


\subsection{$p^\uparrow\, p \,\to D\,X$}
At LO  $D$ mesons can be produced from the fragmentation of a $c$ or $\bar c$ quark created  either through the hard  subprocess $q\bar q \to c\bar c$, or through $gg\to c\bar c$. In terms of the invariants
\begin{equation}
\tilde s  \equiv (p_a+p_b)^2 = \hat s\,,\qquad   \tilde{t} \equiv (p_a-p_c)^2-m_c^2= \hat t -m_c^2\,,\qquad \tilde{u} \equiv  (p_b-p_c)^2-m_c^2 = \hat u -m_c^2\,,
\end{equation}
where $m_c$ is the charm mass, the CGI-GPM hard functions for the gluon fusion processes read 
\begin{align}
H^{{\rm Inc} \,(f)}_{gg\to c\bar c} & = H^{{\rm Inc}\,(f)}_{gg\to \bar c c} = -\frac{N_c}{4(N_c^2-1)}\,  \frac{1}{\tilde t \tilde u} \, \left (  \frac{\tilde t^2}{\tilde s^2}+ \frac{1}{N_c^2}\right ) \left ( \tilde t^2 + \tilde u^2 + 4 m_c^2 \tilde s - \frac{4 m_c^4 \tilde s^2}{\tilde t \tilde u }    \right )\, ,\nonumber \\
H^{{\rm Inc} \,(d)}_{gg\to c\bar c} & =- H^{{\rm Inc}\,(d)}_{gg\to \bar c c} = -\frac{N_c}{4(N_c^2-1)}\,  \frac{1}{\tilde t \tilde u} \, \left (  \frac{\tilde t^2 - 2 \tilde u^2}{\tilde s^2}+ \frac{1}{N_c^2}\right ) \left ( \tilde t^2 + \tilde u^2 + 4 m_c^2 \tilde s - \frac{4 m_c^4 \tilde s^2}{\tilde t \tilde u }    \right )\,.
\label{eq:Hfd}
\end{align}
These results have been obtained using the same method employed previously for $p^\uparrow p \to J/\psi \, X$ and are in agreement with the ones in Ref.~\cite{Kang:2008ih}, calculated in the collinear twist-three approach. Moreover, for the light quark annihilation we find
\begin{align}
H^{{\rm Inc}}_{q\bar q\to  \bar c  c}  = - H^{{\rm Inc}}_{\bar q q \to   c \bar c} = \frac{3}{N_c^2-1}\,H^{{\rm Inc}}_{q\bar q\to c\bar c}= -  \frac{3}{N_c^2-1}\,H^{{\rm Inc}}_{\bar q q\to \bar c c} =  \frac{3}{2N_c^2}\,\left (  \frac{\tilde t^2 + \tilde u^2 + 2 m_c^2\tilde s}{\tilde s^2} \right ) \,.
\label{eq:Hq}
\end{align}
As pointed out in Ref.~\cite{DAlesio:2017rzj}, all the contributions to $A_N$, other than the Sivers one, can be safely neglected because they enter with azimuthal phase factors that strongly suppress them after integration over transverse momenta.

\subsection{$p^\uparrow p\to \pi\, X $ }
\label{pion}

The hard functions to be convoluted with the gluon Sivers distribution $f_{1T}^{\perp g\, (f)}$ are given by~\cite{DAlesio:2018rnv}
\begin{align}
&H_{gq\to gq}^{\text{Inc} \,(f)} = H_{g\bar{q}\to g\bar{q}}^{\text{Inc} \,(f)}  = -\frac{\hat s^2+\hat u^2}{4 \hat s \hat u}\, \bigg (  \frac{\hat s^2}{\hat t^2} + \frac{1}{N_c^2} \bigg ) \,,\nonumber\\
& H_{gq\to qg}^{\text{Inc} \,(f)} = H_{g\bar{q}\to \bar{q} g}^{\text{Inc} \,(f)}  = -\frac{\hat s^4-\hat t^4}{4 \hat s \hat t\hat u^2} \,, \nonumber     \\
&H_{gg\to q \bar q}^{\text{Inc} \,(f)}  = H_{gg\to \bar{q} q}^{\text{Inc} \,(f)}=-\frac{N_c}{4 (N_c^2-1)}\, \frac{\hat t ^2+ \hat u^2}{ \hat t \hat u} \,\bigg (\frac{\hat t^2}{\hat s^2}\, +\frac{1}{ N_c^2} \bigg )\,,\nonumber \\
&H_{gg\to gg}^{\text{Inc} \,(f)} = \frac{N_c^2}{N_c^2-1}\,\bigg (\frac{\hat t}{\hat u} - \frac{\hat s}{\hat u}\bigg)\, \frac{(\hat s^2 +  \hat s \hat t + \hat t^2)^2}{\hat s^2 \hat t^2}\, ,
\end{align}
while for the other gluon Sivers function, $ f_{1T}^{\perp g\, (d)}$, one has
\begin{align}
&H_{gq\to gq}^{\text{Inc} \,(d)} = -H_{g\bar{q}\to g\bar{q}}^{\text{Inc} \,(d)}= \frac{\hat s^2 + \hat u^2}{ 4 \hat s \hat u} \, \bigg ( \frac{ \hat s^2-2\hat u^2}{\hat t^2} + \frac{1}{N_c^2}  \bigg ) \,,\label{eq:Hgqgq}\\
& H_{gq\to q g}^{\text{Inc} \,(d)} = -H_{g\bar{q}\to \bar{q} g}^{\text{Inc} \,(d)} = -\frac{\hat s^2 + \hat t^2}{4 \hat s \hat t } \, \left (\frac{\hat s^2 + \hat t^2}{\hat u^2}  - \frac{2}{N_c^2}    \right )  \,,\label{eq:Hgqqg}\\
& H_{gg\to q \bar q}^{\text{Inc} \,(d)}  = -H_{gg\to \bar{q} q}^{\text{Inc} \,(d)} =-\frac{N_c}{4 (N_c^2-1)}\, \frac{\hat t ^2+ \hat u^2}{ \hat t \hat u} \,\bigg ( \frac{\hat t^2-2 \hat u^2}{\hat s^2} + \frac{1}{N_c^2} \bigg )\,,\\
& H_{gg\to gg}^{\text{Inc} \,(d)} = 0\,\label{eq:Hgggg}.
\end{align}
We can therefore conclude that, in the CGI-GPM,  both gluon Sivers functions contribute to the asymmetry. In addition,  one needs to take into account the effect due to the quark Sivers function as well, which has been calculated in Ref.~\cite{Gamberg:2010tj}. Furthermore, in the case of pion production,  $A_N$ can in principle arise from the competing Collins mechanism, due to the fragmentation of transversely polarized quarks~\cite{Collins:1992kk}. The Collins effect is however suppressed at central rapidities for RHIC kinematics~\cite{DAlesio:2018rnv}. 

\subsection{$p^\uparrow p \to \gamma\, X$}
\label{photon}

The modified partonic hard functions for the gluon induced subprocesses read~\cite{DAlesio:2018rnv}
\begin{align}
& H^{\text{Inc}\,(f)}_{gq\to \gamma q} =  H^{\text{Inc}\,(f)}_{g\bar{q}\to \gamma \bar{q}} = -\frac{1}{2} \,H^U_{gq\to \gamma q} 
\, , \label{eq:Hgqgammaq1}\qquad\quad H^{\text{Inc}\,(d)}_{gq\to \gamma q} = -H^{\text{Inc}\,(d)}_{g\bar{q}\to \gamma \bar{q}} = \frac{1}{2} \,H^U_{gq\to \gamma q} \,,
\end{align}
while the (anti)quark induced ones can be found in Ref.~\cite{Gamberg:2010tj}.

\section{Concluding remarks}

The results presented here can be used to put preliminary constraints on the two independent gluon Sivers functions $ f_{1T}^{\perp g\, (f)}$ and $ f_{1T}^{\perp g\, (d)}$. This is achieved by performing a combined phenomenological analysis of existing RHIC data on SSAs for $p^\uparrow  p\to \pi\, X$ and $p^\uparrow p\to D\, X$. The outcome of this study, which  is summarized in a follow-up to this contribution~\cite{DAlesio:2019}, is used to provide predictions for the process $p^\uparrow p\to J/\psi\,X$, for which measurements are available, and $p^\uparrow  p\to \gamma\, X$, currently under investigation at RHIC. 
It turns out that  present  data do not allow for a clear discrimination between the CGI-GPM and the simpler GPM approach, which assumes  a single universal gluon Sivers function and does not take into account  initial and final state interactions~\cite{DAlesio:2018rnv,DAlesio:2019}.

\end{document}